# Two-dimensional electron systems in perovskite oxide heterostructures: Role of the polarity-induced substitutional defects

Shih-Chieh Lin,[1,2] Cheng-Tai Kuo,[1,2,3,*] Yu-Cheng Shao,[4] Yi-De Chuang,[4] Jaap Geessinck,[5] Mark Huijben,[5] Jean-Pascal Rueff,[6,7] Ismael L. Graff,[8] Giuseppina Conti,[1,2] Yingying Peng,[9,#] Aaron Bostwick,[4] Eli Rotenberg,[4] Eric Gullikson,[2] Slavomír Nemšák,[4] Arturas Vailionis,[10,11] Nicolas Gauquelin,[5,12] Johan Verbeeck,[12] Giacomo Ghiringhelli,[9] Claus M. Schneider,[1,13] and Charles S. Fadley[1,2,†]

[1]Department of Physics, University of California Davis, Davis, California 95616, USA
[2]Materials Sciences Division, Lawrence Berkeley National Laboratory, Berkeley, California 94720, USA
[3]Stanford Synchrotron Radiation Lightsource, SLAC National Accelerator Laboratory, Menlo Park, California 94025, USA
[4]Advanced Light Source, Lawrence Berkeley National Laboratory, Berkeley, California 94720, USA
[5]Faculty of Science and Technology and MESA+ Institute for Nanotechnology, University of Twente, Enschede 7500 AE, The Netherlands
[6]Synchrotron SOLEIL, L'Orme des Merisiers, Saint-Aubin-BP48, 91192 Gif-sur-Yvette, France
[7]Sorbonne Université, CNRS, Laboratoire de Chimie Physique-Matière et Rayonnement, 75005 Paris, France
[8]Department of Physics, Federal University of Paraná, Curitiba, Brazil
[9]CNR-SPIN and Dipartimento di Fisica Politecnico di Milano, Piazza Leonardo da Vinci 32, Milano I-20133, Italy
[10] Stanford Nano Shared Facilities, Stanford University, Stanford, California 94305, USA
[11]Department of Physics, Kaunas University of Technology, Studentu street 50, LT-51368 Kaunas, Lithuania
[12] Electron Microscopy for Materials Science (EMAT), University of Antwerp, Groenenborgerlaan 171, B-2020 Antwerp, Belgium
[13]Peter-Grünberg-Institut PGI-6, Forschungszentrum Jülich, Jülich 52425, Germany
* Email: ctkuo@slac.stanford.edu
# Present address: International Center for Quantum Materials, School of Physics, Peking University, Beijing 100871, China
† Deceased August 1, 2019.

## ABSTRACT

The discovery of a two-dimensional electron system (2DES) at the interfaces of perovskite oxides such as $LaAlO_3$ and $SrTiO_3$ has motivated enormous efforts in engineering interfacial functionalities with this type of oxide heterostructures. However, the fundamental origins of the 2DES are still not understood, e.g. the microscopic mechanisms of coexisting interface conductivity and magnetism. Here we report a comprehensive spectroscopic investigation on the depth profile of 2DES-relevant Ti 3d interface carriers using depth- and element-specific techniques like standing-wave excited photoemission and resonant inelastic scattering. We found that one type of Ti 3d interface carriers, which give rise to the 2DES are located within 3 unit cells from the n-type interface in the $SrTiO_3$ layer. Unexpectedly, another type of interface carriers, which are polarity-induced Ti-on-Al antisite defects, reside in the first 3 unit cells of the opposing LaAlO$_3$ layer (~10 Å). Our findings provide a microscopic picture of how the localized and mobile Ti 3d interface carriers distribute across the interface and suggest that the 2DES and 2D magnetism at the LaAlO$_3$/SrTiO$_3$ interface have disparate explanations as originating from different types of interface carriers.

**I. INTRODUCTION**

The discovery of conductivity at the polar-nonpolar interfaces of insulating oxides such as LaAlO$_3$ (LAO) and SrTiO$_3$ (STO) has revealed great potentials for engineering emergent interfacial functionalities absent in their bulk forms [1,2,3,4,5]. The two band insulators LAO and STO have the perovskite structure consisting of a mutual stacking of (LaO)$^+$ and (AlO$_2$)$^-$, and (SrO)$^0$ and (TiO$_2$)$^0$ atomic layers, respectively, with their nominal valence values indicated. In the [001] direction, two different interfaces can be formed between the polar LAO and nonpolar STO: (LaO)$^+$/(TiO$_2$)$^0$ (named n-type) and (SrO)$^0$/(AlO$_2$)$^-$ (named p-type) [1,2,3]. A remarkable feature is that the two-dimensional electron system (2DES) can only form at the n-type interface, when the thickness of the top LAO is beyond a critical thickness of 3 unit cells (uc) [3].

How is this 2DES established in the first place? While there is some consensus that its origin may be associated with the Ti 3d electrons at the interfaces [6], the underlying microscopic mechanisms are far from being understood. A variety of suggestions have been put forward to explain some of the experimental observations, but they usually address only specific aspects and are yet unable to provide a consistent solution to all the puzzles. For example, the mechanism of intrinsic electronic reconstruction [1,2,3,7] and mechanisms that involve oxygen vacancies [8,9,10,11] and interfacial cation intermixing [12,13,14,15,16] are able to interpret the interface conductivity at n-type interfaces, but they are difficult to explain the insulating p-type interfaces and the interface magnetism. In 2014 Yu and Zunger took a very interesting approach and went beyond the view of a mere electronic reconstruction. On the basis of first principles calculations, they proposed the so-called polarity-induced defect mechanism [17]. It predicts that defects spontaneously form at the LAO surface and/or the LAO/STO interfaces in order to compensate the built-in field induced by the polar-discontinuity, and in this way avoid a divergence to the electric potential. Consequently, the formation of 2DES and the appearance of interface magnetism ought to be driven by the polar-field-induced defects, such as the paired antisite and oxygen vacancy defects [17]. Exploring such defects in experiments is a challenging task and calls for sophisticated approaches that are able to extract information specifically from the interface.

X-ray photoemission spectroscopy is a powerful tool for revealing the interface electronic structure and has provided valuable information regarding the electronic reconstruction and quantum confinement effect of these occupied Ti 3d states that contribute to 2DES [18,19,20,21,22,23]. More specifically, resonant X-ray photoemission spectroscopy (RXPS) can enhance the spectral weight of Ti 3d states around the Fermi level position and thus has been widely used to study the interface electronic structure of LAO/STO heterostructures [20,21,22,23]. By scanning the photon energies across the Ti 2p core level, RXPS can distinguish if the interface carriers are associated with the in-gap (IG) states resulting from oxygen vacancies or the quasiparticle (QP) states from 2DES [23,24]. Complementary information about the Ti electronic states can be obtained from resonant inelastic X-ray scattering (RIXS). It probes Ti 3d orbital transitions, namely *dd* excitations, between occupied and unoccupied states [25,26,27,28]. As a drawback, neither conventional photoemission nor RIXS are able to probe the depth profile of the emitted photoelectrons or scattering photons across an interface. Such a profile is strongly needed, however, to evaluate the real space distribution of the Ti 3d carriers in the LAO/STO interface. By contrast, cross-sectional scanning probe microscopies (XSPM) and scanning transmission electron microscopy (STEM) offer high spatial resolution and have been widely used to study the local electronic properties relative to the 2DES of LAO/STO heterostructures [12,29,30,31,32], but they cannot resolve the electronic/orbital states. Some advanced STEM studies were able to localize electronic states and hybridization [33] and image orbitals [34] but they are limited in the determination of orbital character due to low q-resolution [33] and subject to instrumental effects and delocalization of the inelastic scattering [35,36].

We have demonstrated in the prior works that standing wave-excited (SW) spectroscopic techniques, such as SW-RXPS, SW hard X-ray photoemission spectroscopy (SW-HXPS) [37,38,39,40,41], and SW-RIXS [42], are capable of disentangling the physics at buried interfaces in terms of depth profiling the elemental diffusions, polarization-induced voltage drop, and orbital and magnetic excitations. They are thus uniquely suited to elucidate the formation of the 2DES at the LAO/STO interface. In this work, we have therefore used a full suite of SW-excited spectroscopic approaches to extract the depth profile of the orbital character of Ti 3d interface electrons across the LAO/STO interfaces and to detail its influence on the interface conductivity and magnetism.

**II. Results**

**A. Depth sensitivity of X-ray standing wave techniques**

Figure 1(a) shows a schematic of SW measurements for this specific sample configuration. The superlattice sample consists of 20 bilayers of 5 uc LAO and 5 uc STO layers on a TiO$_2$-terminated STO (001) substrate. Such limited thicknesses for both layers provide a good basis for visualizing how the 2DES, and possible defects, distribute across the LAO/STO interfaces. Furthermore, the layer thicknesses are chosen such that the superlattice will exhibit metallic behavior with high mobility charge carriers, similar to a single LAO/STO interface system. In a previous study we have shown that adding an additional STO layer on top of the LAO layer triggers the electronic conductivity at a significantly lower LAO film thickness than for the uncapped systems [2], and even a 2 uc STO capping layer is enough to result in very similar metallic behavior as a 10 uc STO capping layer [7], see Supplemental Fig. 2 for the typical metallic behavior in our LAO/STO superlattice [43].

The strain condition in our LAO/STO superlattice is equal to the previous studies on LAO/STO interfaces in various sample architectures (single film, bilayer, multilayer or superlattice) on top of single crystalline STO substrates, in which the LAO and STO layers are fully in-plane strained by the underlying STO substrate (3.905 Å) [10]. The STEM characterization in Supplemental Fig. 3 [43] shows the controlled epitaxial ordering throughout the full superlattice with a constant in-plane lattice parameter. The out-of-plane lattice parameters of both layers were determined from the peak positions in the XRD analysis, as shown in Supplemental Fig. 1, to be ~3.73 Å and ~3.90 Å for the respective LAO and STO layers. Our superlattice shows the typical shortening of the out-of-plane

LAO lattice parameter (instead of 3.791 Å seen in the bulk), while the STO layers remain strain-free [10].

To better demonstrate the difference in the depth sensitivity for these SW techniques, their calculated yield strength distributions are shown in Fig. 1(b)-(d). Figure 1(b) and (c) are the SW-HXPS ($hv$ = 3000 eV) and SW-RXPS ($hv$ = 459 eV) photoemission yield distributions while Fig. 1(d) is the SW-RIXS ($hv$ = 459.3 eV) yield distribution. The probing depth for RXPS, HXPS, and RIXS determined from the yield distributions is ~27, 135, and 960 Å, respectively (see Methods), and the shorter depth for RXPS and HXPS relative to RIXS is due to the shorter escape depth for photoelectrons. Hence, SW-RXPS is sensitive to the top LAO layer and the first interface, SW-HXPS is sensitive to the top 3 LAO/STO (STO/LAO) interfaces, and SW-RIXS is sensitive to all 20 LAO/STO (STO/LAO) interfaces. We have combined these different depth sensitivities to obtain a more consistent picture.

Although RIXS has the longest probing depth among these SW techniques, its signal comes only from the Ti ions in the STO layer, a partial information that cannot be used to derive the full sample structure as done with SW photoemission. Such short-fall can be circumvented by using the core-level rocking curves (RCs) from both SW-HXPS and SW-RXPS to better determine the multilayer structure. We note that the information obtained from HXPS is particularly useful for determining the depth profile of near Fermi (NF) peak measured with SW-RXPS and also the *dd* excitations from SW-RIXS.

**B. Standing-wave excited photoemission results**

Figures 2(a)&(b) show the experimental core-level spectra and their corresponding RCs (open circles) from representative La and Sr core-levels at $hv = 3000$ eV and at $hv = 459$ eV, respectively. The best-fit simulated core-level RCs (curves) are plotted together with the experimental RCs. The best-fit superlattice sample structure was determined by minimizing the total difference between 8 experimental (both soft and hard X-rays) RCs and the simulated RCs simultaneously via iteratively adjusting the input structure. The whole fitting results regarding the core-level spectra and their RCs are shown in Supplemental Fig. 6&7&8[43]. The determination of best-fit RCs is described in the Methods section. The best-fit results indicate that the superlattice sample has reasonably sharp interface. The averaged interdiffusion between the interfaces is 7 Å with respect to the ideal sample and is in excellent agreement with the STEM results of 1~3 uc elemental diffusions (see Supplemental Fig. 3 [43])

Next, we focus on determining the depth profile of the Ti 3d electronic states using SW-RXPS. Using RXPS with excitation energies near the Ti $L_3$ edge, the spectral weight of Ti 3d states around the Fermi level position in the valence band spectrum of LAO/STO heterostructures can be enhanced [20,21,22,23,24]. As mentioned earlier, QP and IG states can be observed around the Fermi level position while excited at the QP and IG resonances. The photon-energy-scanned RXPS map in Supplemental Fig. 5 shows that the chosen energy for SW-RXPS resonance is in the middle of the QP and IG resonances. Due to the limited RXPS energy resolution and the chosen SW resonance, the QP and IG peaks merge into one peak whose centroid shifts to a higher energy. This specific SW-RXPS energy was selected for maximizing the reflectivity/standing wave effect, but apparently it causes the drawback of indistinguishable IG and QP contributions. A detailed discussion can be found in the Supplemental Note 3 [43].

Figure 2(c) shows the RXPS valence band spectrum with an inset for a magnified view around the Fermi level where a near Fermi (NF) peak centered at ~0.35 eV is observed. The slightly higher NF peak position with respect to the prior work [20] can be related to the limited energy resolution, and the chosen SW resonance, or from a different level of the impurities in different samples. This NF peak includes both contributions from QS and IG states. The experimental NF (open circles) and its best-fit (curve) RCs are shown in Fig. 2(b). The contrasting NF RC relative to other core-level RCs indicates that this NF component has its own depth distribution. Since NF is associated with the Ti 3d orbitals, its depth distribution can be related to the Ti 3d interface carriers, or the 2DES in the STO layer [20,23,24]. From the depth profile at the first LAO/STO interface (see Fig. 2(d)), our SW-RXPS results suggest that an unexpected source of Ti 3d interface carriers with high concentration is observed in the LAO layer, which will be discussed later.

## C. Standing-wave excited resonant inelastic X-ray scattering results

To gain further insight on the depth distribution of Ti 3d orbitals and its relationship to the 2DES formation, we have performed SW-RIXS measurements on the very same sample and the results are summarized in Figure 3. In order to distinguish the Raman-like excitations from the normal fluorescence emission, the excitation-energy-dependent RIXS measurements were performed near the Ti $L_3$ edge, see Fig. 3(a) (the wider range energy loss spectrum showing the charge-transfer component can be found in Supplemental Fig. 10 [43]). The *dd* component shows a constant Raman shift of ~2.6 eV, whereas the fluorescence has a linear shift with excitation photon energy. The chosen excitation energy for the SW-RIXS measurements was 1 eV below the $e_g$ resonance to better assist probing the $Ti^{3+}$ ions and separating the *dd* and fluorescence features. Based on the previous work, we can fit the SW-RIXS spectra with five peaks: quasi-elastic line, $d_{xz}/d_{yz}$ (0.8 eV), $d_{x^2-y^2}$ (2.7eV), $d_{z^2}$ excitations (3.2 eV), and fluorescence (~1.8 eV), as demonstrated in Fig. 3(b) [25,27].

Their respective experimental RCs (open circles) and best-fit results (curves) are shown in Figure 3(c). To first order, we can anticipate that the *dd* excitations are coming from Ti 3d interface carriers with associated structural defects and the fluorescence is from the true 2DES phase. The intensity modulation around the Bragg angle (~20°) for these RCs are evidently different and they also do not fit into the simulated RC from the whole STO layer. Consequently, RIXS excitations must have different spatial distributions. Figure 3(d) shows the determined depth distributions in the top 2.5 LAO/STO bilayers for the RIXS excitations. The depth profile for each excitation in Fig. 3(c) was normalized in a way that its integrated intensity is proportional to the angular-averaged RIXS intensity of the corresponding excitations; therefore, the intensity of the depth profile represents a sort of "concentration" map of excitations in depth.

The SW-RIXS results show that there is a notably large amount of Ti 3*d* interface carriers residing in both the LAO and STO sides of the interface, which is consistent with the SW-RXPS results. These interface carriers reside at the n-type interfaces (LAO_bottom/STO_top) and show no contributions at the p-type interfaces (STO_bottom/LAO_top), agreeing with experimental observations of interface conductivity [1]. In general, a RC with larger (smaller) intensity modulation means a more localized (delocalized) distribution in the depth. The intensity modulations in Fig. 3(c) imply that the fluorescence originate from a wider distribution in depth, while the $d_{x^2-y^2}$ excitations originate from narrower distributions. The possible cause of the variations of depth distributions in *dd* excitations will be discussed later.

### III. DISCUSSION AND CONCLUSION

The determined depth profile of the NF peak from SW-RXPS and the *dd* excitations from SW-RIXS represent the depth profiles of Ti 3d interface carriers. These results show that, in contrast to the general assumption that these interface carriers are attributed to the mobile 2DES in the STO layer, there are two different sources residing in both LAO and STO. What does this phenomenon stand for? What is the relationship between these interface carriers and the 2DES?

In Figure 4, we illustrate the polarity-induced defect mechanism for the 2DES formation [17]. The polarity discontinuity across the LAO/STO interfaces triggers the spontaneous formation of multiple kinds of defects. In addition to the surface oxygen vacancy defects ($V_{O(S)}$), the paired antisite defects like Ti-on-Al ($Ti_{Al}$) and Al-on-Ti ($Al_{Ti}$) from the hopping of Ti atoms near the interface into the $AlO_2$ atomic layers and exchanging site with Al atoms, can be present in order to alleviate the polarization-induced field across the interfaces. At this n-type interface, when the LAO thickness is below the critical thickness, all electrons are transferred from $Ti_{Al}$ to $Al_{Ti}$ defects with lowest formation energy to cancel the polar field and subsequently these electrons will be trapped by deep $Al_{Ti}$ defects in the STO layer; as a consequence, these electrons become immobile. On the other hand, when the LAO thickness is above the critical thickness, the $V_{O(S)}$ defects are responsible for the cancelation of the built-in polar field [17]. Among the electrons ionized from the $V_{O(S)}$ defects, half of them are donated to the $Al_{Ti}$ defects and the rest form the 2DES in the STO layer. Meanwhile, the interface $Ti_{Al}$ defects in the LAO layer ($Ti^{3+}$ ions on $Al^{3+}$ sites) donate no electrons. Some of the surface $Ti_{Al}$ defects still tend to donate electrons and some remain ionized when the LAO thickness is larger than the critical thickness, but it will gradually be dominated by the $V_{O(S)}$ defects with the increased LAO thickness. These polarity-induced substitutional defects in LAO can be probed by RXPS and RIXS, in addition to the 2DES.

The Ti ions in the STO layer can be regarded as the mobile 2DES, residing within the 3 uc STO layer (~12 Å in length) near the interface, of which the length scale is fairly consistent with the recent STEM results of ~10±03 Å [32]. The other Ti ions located at LAO layer can be the $Ti_{Al}$ defects, residing within ~10 Å LAO layer near the interface. This length scale also perfectly matches to the critical thickness of 3 uc predicted by the first principles calculations [17]. Note that a small concentration of $Ti_{Al}$ defects was also found at the LAO surface because some of the $Ti_{Al}$ defects are ionized as $Ti^{3+}$ at the LAO surface; this is also consistent to the polarity-induced defect mechanism [17]. These findings show the strong experimental evidence that substantiates the polarity-induced defect mechanism.

The depth profiles of the NF and *dd* excitations all show the coexistence of the polarity-induced defects ($Ti_{Al}$) and 2DES. Additionally, the $Ti_{Al}$ defects and 2DES have identical NF peak and energy levels, meaning that they both exhibit orbital reconstructions and quantum confinement effects. This phenomenon implies a strong correlation between these $Ti_{Al}$ defects and mobile 2DES. However, the nature of these two Ti interface carriers are still different in terms of the mobility. The polarity-induced defects are localized across the interfaces while the 2DES is mobile. As discovered by the previous HXPS and RIXS results, the derived sheet carrier densities were significantly higher than those obtained by Hall effect measurements, suggesting the coexistence of localized and mobile Ti $3d$ interface carriers [18,25]. In these works, RIXS/HXPS were utilized for the detection of $Ti^{3+}$ carriers; they found that an earlier onset (2 uc) of observing Ti 3d interface carriers by RIXS and HXPS compared to that by Hall effect results (4 uc). The quantity of the RIXS/HXPS-determined interface carriers is much higher than that of the Hall-effect-determined 2DES. Both RIXS and HXPS techniques are not able to tell whether these interface carriers are localized or mobile and nor to determine their spatial distribution. Although this phenomenon provides a hint that these RIXS/HXPS-detected carriers are not merely 2DES in the STO layer, the microscopic mechanism is still unclear. Our work provides a reasonable interpretation of this phenomenon. Instead of only observing the 2DES, the RXPS/RIXS experiments observe both the defects and 2DES, because they have $Ti^{3+}$ characteristics. In other words, when the LAO thickness is less than 3 uc, RIXS/RXPS detect the polarity-induced defects. When the LAO thickness is above the critical thickness, RIXS/RXPS detect both polarity-induced defects and 2DES.

Our findings further provide a clear physical picture of depth distribution of these two coexisting carriers. It is noteworthy that, according to the polarity-induced defect mechanism [17], these trapped $Ti^{3+}$ ions ($Ti_{Al}$) in LAO layer are responsible for the interface magnetism, while the mobile 2DES in STO layer is responsible for the interface conductivity. The understanding of the interplay of the layers of mobile electrons and trapped ions will be important to enable controlling both interface conductivity and magnetism, which warrants further investigation.

As for the results from SW-RIXS, the variations in the depth distributions of $dd$ excitations in Figure 3(c) might be seen as insignificant to affect the main messages presented in the previous paragraphs. To give a meaning to the differences in the RC of the various $dd$ contributions, the subband picture [19,25,28] sketched in Fig. 4(b) can be utilized. Due to the quantum confinement effect, the half-filled $d_{xy}$ orbital is confined at the bottom of a quantum well and has the narrowest depth extension. In contrast, the $d_{xz}/d_{yz}$ orbitals at higher energy are more weakly confined, and this effect is even stronger for the $d_{x^2-y^2}$ and $d_{z^2}$. Finally, the fluorescence signal shows wider distribution with more contributions in STO layer. Indeed, the differences among the $dd$ excitations depth profiles is an intriguing phenomenon, yet difficult to explain at the qualitative level; a more quantitative interpretation will have to include the RIXS cross section and the quantum confinement which goes beyond the scope of present work.

The precision of determining the depth distribution is limited by the SW wavelength, which is ~3.92 Å in this work (details see Supplemental Note 2 [43]). Therefore, the findings of two kinds of Ti interface carriers residing in the LAO and STO layers are not affected by the energy resolution of RXPS and RIXS. However, the limited energy resolution and the chosen resonance for SW-RXPS leaves a single NF peak without resolving the 2DES and IG states. This means that the depth distribution of 2DES, polarity-induced defects, and oxygen vacancies (IG states) can't be discriminated. The oxygen vacancies are one type of source to the $Ti^{3+}$ carriers and they are not necessarily locating merely at the LAO or STO layer. If improving the energy resolution of RXPS, this technique can definitely provide insights to the depth distribution of the oxygen vacancies and the relationship between the oxygen vacancies, polarity-induced defects, and 2DES.

Another future direction of SW spectroscopic studies on the LAO/STO system is to explore the role of oxygen vacancies. Recently, it is found that the LAO/STO interface electronic structure, which includes the QP (associated with 2DES) and IG states (associated with localized electrons), is tunable through the control of the oxygen vacancies [24,44,45,46]. SW spectroscopies can show the depth profiles of both the IG and QP states in the oxygen deficient LAO/STO heterostructure, revealing the relationship between oxygen vacancies and 2DES. In addition, the determined 2DES distribution can be related to the width of quantum well that confines the 2DES. Prior work [47] demonstrates the control of the 2DES transport properties upon the electrostatic doping. In this case, the quantum well width across the LAO/STO interfaces is tuned by the variation of a gate voltage. Using SW techniques, one can attain more quantitative understanding of quantum well control, which could be another interesting direction for future studies.

We have demonstrated the coexistence of polarity-induced substitutional defects and 2DES across the polar/non-polar oxide interfaces, implying that the spontaneously formed defects for canceling the interfacial polarization field are the fundamental origins of 2DES formation. Our findings indicate that the well-accepted assumption that the Ti 3d interface electrons only exist in the STO side of the interface is an over-simplified picture and should be revised. In addition, the interface conductivity and magnetism at LAO/STO interfaces could originate from different Ti interface carriers located in the STO and LAO layer, respectively. In a methodological perspective, we have demonstrated the potential of combining SW photoemission and RIXS spectroscopies to obtain the depth distribution of electronic/orbital states at buried interfaces with precision of the Angstroms depth precision.

## IV. METHODS

### A. Sample synthesis

The superlattice sample, $[(LaAlO_3)_n/(SrTiO_3)_m]_p$, $(LAO_n/STO_m)_p$ (n= 5 unit cell (uc), m= 5 uc, and p = 20), was deposited by pulsed laser deposition on a $TiO_2$-terminated Nb-doped $SrTiO_3$ (001) substrate. The individual materials were ablated from single-crystalline $LaAlO_3$ and $SrTiO_3$ targets using a KrF excimer laser operating at 248 nm, while laser fluence and repetition rate were 1.3 J/cm$^2$ and 1 Hz, respectively. To enable the formation of 2DES interfaces in the LAO/STO multilayers, an oxygen pressure of $2 \times 10^{-3}$ mbar was applied, while maintaining a growth temperature of 800 ºC [2,3,7]. Reflective high-energy electron diffraction indicated layer-by-layer growth mode during the full superlattice growth and was used to enable the precise unit-cell-control of all individual layers.

### B. Sample characterization

X-ray diffraction (XRD) analysis confirmed the epitaxial growth of the superlattice as it is fully strained in plane to the STO cubic structure. The highly ordered growth of the superlattice is further demonstrated by the observation of clear Laue fringes between the diffraction peaks (see Supplemental Fig. 1(a) [43]). Analysis of the XRD results showed the coherent growth with a reduced c-axis parameter of 3.73 Å for each LAO layer compared to 3.90 Å for each STO layer, which is in good agreement with previous reports [10,48,49]. The low level of surface roughness was confirmed by atomic force microscopy analysis of the surface of LAO/STO superlattice, indicating the presence of smooth terraces separated by clear, single-unit-cell height steps similar to the surface of the initial $TiO_2$-terminated STO (001) substrate (see Supplemental Fig. 1(b) [43]). The structural and chemical quality of the sample have been verified on the local scale by scanning transmission electron microscopy (STEM), high angle annular dark field (HAADF) imaging (Z-Contrast), and electron energy loss spectroscopy (EELS) (See Supplemental Fig. 3 [43]). STEM was performed on a Titan microscope operated at 300keV with a probe semi convergence angle of 21 mrad; EELS measurements were performed collecting the Ti $L_{2,3}$, Sr $L_{2,3}$ edge, O K and La $M_{4,5}$ and Al K edges simultaneously. The acquisition parameters were 0.04 s/pixel, 0.2 Å/pixel and 0.5 eV/pixel for mapping in the dual EELS mode. Collection semi-angles for HAADF imaging and EELS were 41-94 mrad and 94 mrad, respectively. STEM results show the superlattice sample has sharp and well-controlled interfaces with high quality that is compatible with the previous works [12,50].

## C. Standing wave excited hard and resonant X-ray photoemission (SW-HXPS and SW-RXPS) measurements

SW-RXPS measurements were performed at the beamline MAESTRO of Advanced Light Source, Lawrence Berkeley National Laboratory. SW-HXPS measurements were performed at the beamline GALAXIES of SOLEIL synchrotron. The synchrotron radiation light was *p*-polarized. The energy resolution of SW-RXPS is 500 meV and that of SW-HXPS is 440 meV. Both measurements were performed at room temperature to prevent the charging effect. The details of GALAXIES endstation can be found elsewhere [51]. In all the RXPS spectra, the contributions of the 2$^{nd}$ order light were subtracted [24,45] and its details are discussed in Supplemental Materials [43]. The intensities and profiles of the near Fermi level peak were examined to check the X-ray irradiation effect. From the spectra measured before and after a complete set of SW-RXPS measurements (~12 hours), no clear change was observed. Therefore, we conclude that the irradiation effect is below the detection limit presumably due to the low photon flux of X-rays in our experiments.

**D. Standing wave excited resonant inelastic x-ray scattering (SW-RIXS) measurements**

SW-RIXS measurements were performed at the qRIXS endstation at beamline 8.0.1.3 at the Advanced Light Source, Lawrence Berkeley National Laboratory. The combined energy resolution of the SW-RIXS determined from the FWHM of the elastic peak is 360 meV. The multilayer sample was cooled down to ~78K by liquid $N_2$ during the measurements. The RIXS measurements were probed by p-polarized X-ray with a scattering angle of 115° near the Ti $L_3$ edge. The details of qRIXS endstation can be found elsewhere [52].

**E. X-ray optics calculations of SW effect in photoemission and RIXS**

The SW is generated by the interference between the incident and reflected waves, and it satisfies this equation $\lambda = 2d_{BL}\sin\theta_B$, where $\lambda$ is the wavelength of the incident X-rays, $d_{BL}$ is the thickness of the bilayer, and $\theta_B$ is the Bragg angle. The resulting SW electric-field intensity varies sinusoidally with respect to the sample depth and has a periodicity very close to $d_{BL}$. Scanning the incidence angle over the Bragg peak moves the SW vertically by half of its wavelength, and this variation provides the phase-sensitive depth resolution for photoemission and RIXS. The vertical movement of the SW through the sample with changing the incidence angle will thus enhance or reduce photoemission/RIXS signals from different depths, generating what are called rocking curves (RC). The introduction to the SW technique can be found in the Supplemental Materials.

An X-ray optics computing code (Yang X-ray Optics, YXRO [39]) has been used for generating the yield intensity map and analyzing the experimental RC data. The probing depth for SW photoemission is defined as $3\lambda$, where $\lambda$ is the effective inelastic mean free paths. The probing depth for SW-RIXS is defined as $3\Lambda$, where $\Lambda$ is the effective attenuation lengths. The yield intensities are SW intensities multiplied by an attenuation term which takes into account $\lambda_{RXPS}\backslash\lambda_{HXPS}$ for the photoelectron peaks or $\Lambda$ for the RIXS excitations peaks. The optical constants used for the simulations with excitation energies near the Ti $L_3$ edge were calculated from a Ti XAS spectrum using the Kramers-Kronig relations.

The core-level RCs measured by SW-HXPS and SW-RXPS represent the intensities of photoelectrons collected from the elements distributed uniformly in either the LAO or STO layer. Their best-fit RCs are determined by minimizing the errors between the experimental and simulated RCs simultaneously via iteratively adjusting the input structure (e.g. the LAO thickness). The NF RC for SW-RXPS and *dd* and fluorescence RCs for SW-RIXS don't have uniform distribution and thus have depth profiles with a weighting coefficient for each point at a specific depth. For these kinds of experimental RCs, their best-fit RCs are determined by minimizing the errors between the experimental and simulated RCs simultaneously via optimizing the weighting coefficients (BFGS method). As an example, we choose the NF RC to show the difference between the experimental and simulated RCs using the optimized (best-fit), step-like ($d_{delta}$=1~15 Å), and uniform ($d_{delta}$= whole STO layer) distributions. The optimized distributions show the best results in terms of minimizing the errors. More discussions on determining the depth profiles can be found elsewhere [42].

**ACKNOWLEDGMENTS**

We thank Gabriella Maria De Luca and Lucio Braicovich for discussions. Charles S. Fadley was deceased on August 1, 2019. We thank his significant contributions to this work. We thank Advanced Light Source for the access to Beamline 8.0.3 (qRIXS) via Proposal No. 09892 and beamline 7.0.2 (MAESTRO) via Proposal No. RA-00291 that contributed to the results presented here. We thank synchrotron SOLEIL (via Proposal No. 99180118) for the access to Beamline GALAXIES. This work was supported by the US Department of Energy under Contract No. DE-AC02-05CH11231 (Advanced Light Source), and by DOE Contract No. DE-SC0014697 through the University of California, Davis (S.-C.L., C.-T.K, and C.S.F.), and from the Jülich Research Center, Peter Grünberg Institute, PGI-6. I. L. G. wishes to thank Brazilian scientific agencies CNPQ (Project No. 200789/2017-1) and CAPES (CAPES-PrInt-UFPR) for their financial support. J.V. and N.G. acknowledge funding from the Geconcentreerde Onderzoekacties (GOA) project "Solarpaint" of the University of Antwerp and the European Union's horizon 2020 research and innovation program ESTEEM3 under grant agreement n°823717. The Qu-Ant-EM microscope used in this study was partly funded by the Hercules fund from the Flemish Government.

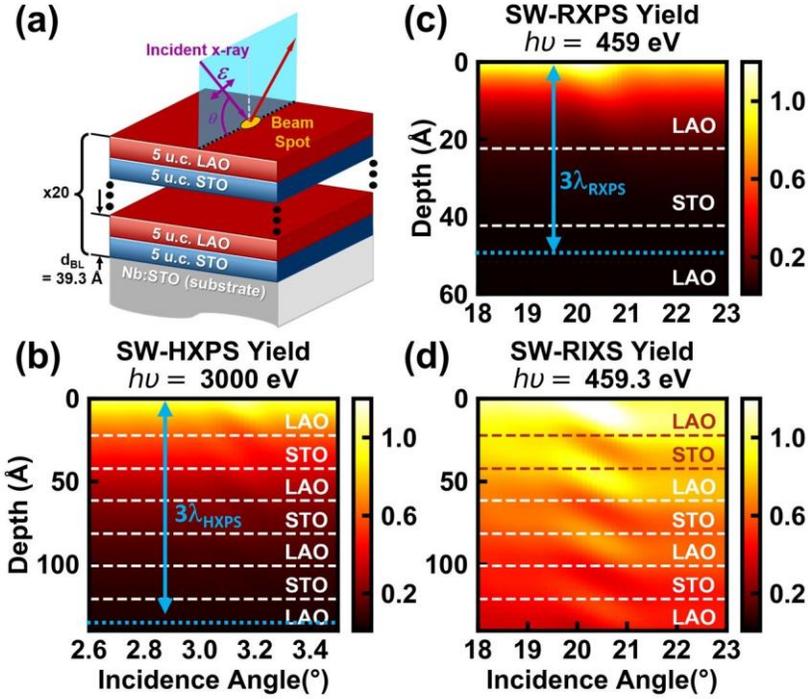

FIG. 1. Depth sensitivity of SW techniques for the LAO/STO superlattice. (a) The superlattice sample, $(LAO_n/STO_m)_p$ (n= 5 uc, m= 5 uc, and p = 20), was used for the SW measurements by varying the X-ray incidence angles $\theta$ around the Bragg angle $\theta_B$. The photoemission yield distribution of (b) SW-HXPS ($hv = 3000$ eV) and (c) SW-RXPS ($hv = 459$ eV). (d) The RIXS yield distribution of SW-RIXS ($hv = 459.3$ eV). The yield distribution demonstrates the probing depth sensitivity of these techniques. The probing depth for RXPS and HXPS ($3\lambda$) is ~27 and 135 Å, respectively, as indicated in b and c. The probing depth for RIXS ($3\Lambda$) is 960 Å (not indicated in d). According to the probing depths, SW-RXPS is sensitive to the top LAO layer and the first interface, SW-HXPS is sensitive to the top 3 LAO/STO (STO/LAO) interfaces, and SW-RIXS is sensitive to all 20 LAO/STO (STO/LAO) interfaces.

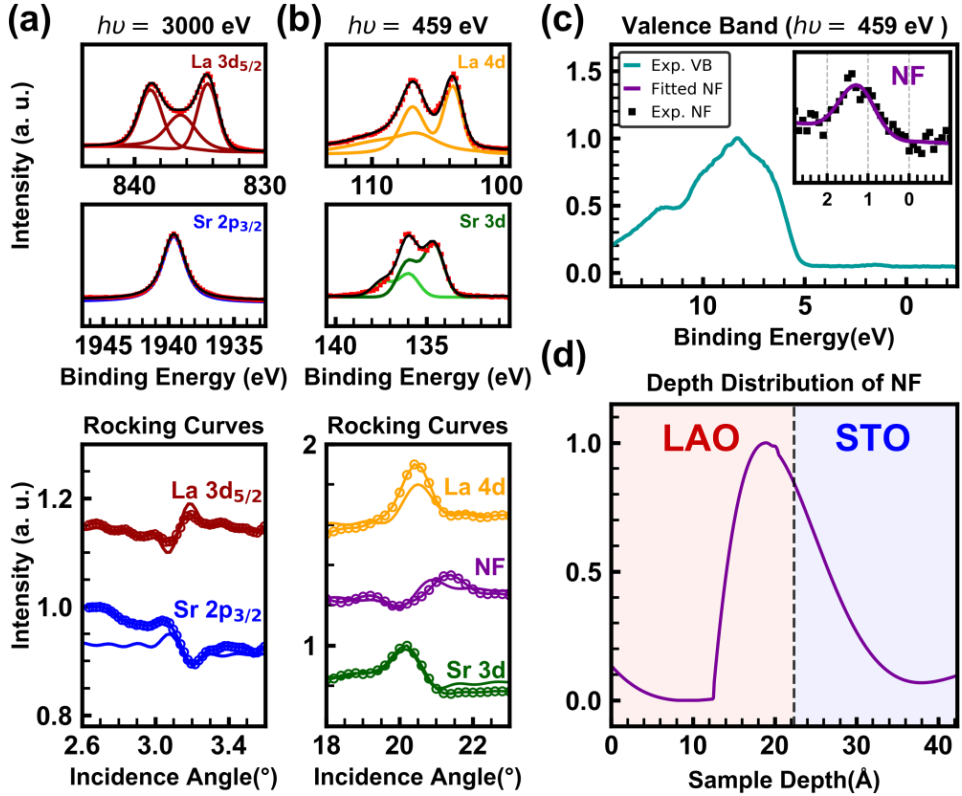

FIG. 2. SW photoemission results and the depth distribution of the near Fermi (NF) peak. (a) SW-HXPS results ($hv = 3000$ eV) (b)&(c)&(d) SW-RXPS results ($hv = 459$ eV). (a) The top panel is the fitted core-level spectra of La $3d_{5/2}$ and Sr $2p_{3/2}$ at an off-Bragg angle. The bottom panel shows their experimental (open circle) and best-fit (curve) rocking curves. (b) The top panel is the fitted core-level spectra of La 4d and Sr 3d. The bottom panel shows the experimental and theoretical rocking curves. For the NF peak, its spectrum is shown in c. (c) A valence band spectrum with an inset of the fitted NF peak near the Fermi level. The NF peak is fitted using Voigt function and Fermi-Dirac distribution. (d) The determined depth profile of the NF peak in the top LAO/STO bilayer. The sample depth of zero means the sample surface.

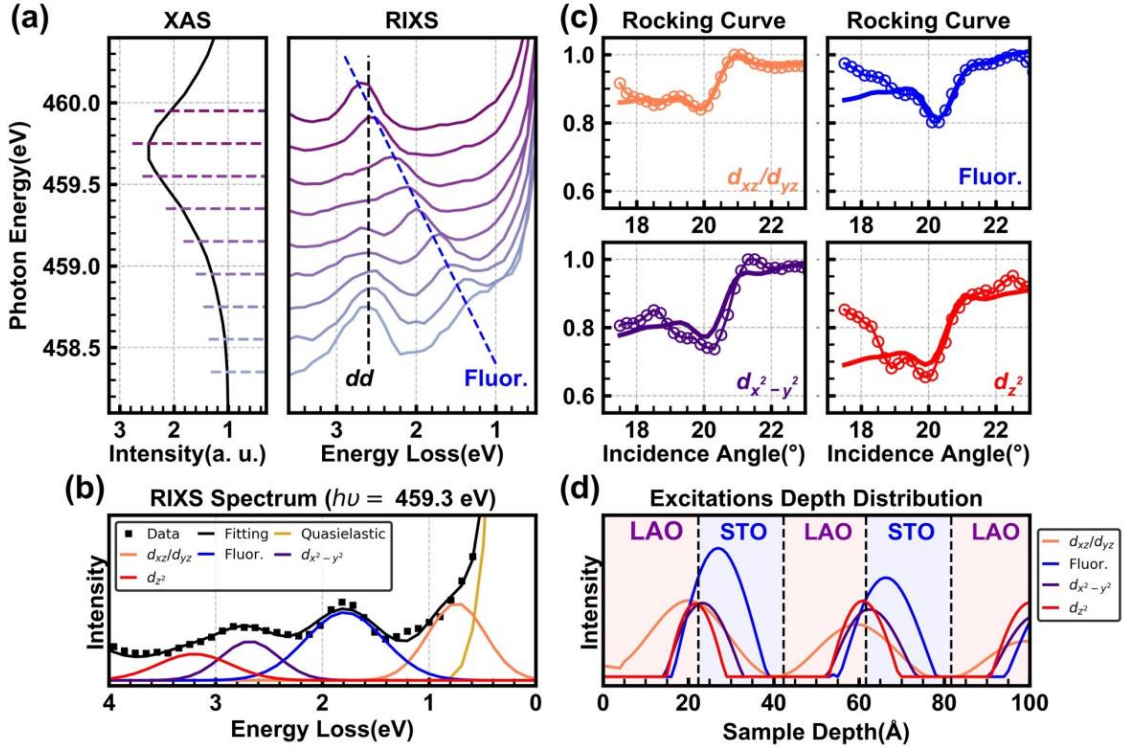

FIG. 3. SW-RIXS result and depth distribution of *dd* excitations. (a) The excitation-energy-dependent RIXS spectra were collected with excitation energies around the Ti $L_3$ XAS edge, ranging from 459.7 to 458.1 eV and marked by color dashed lines in the XAS panel. In the RIXS panel, the black and blue dashed lines are guides to the eyes to show the evolution of the Raman (black) and fluorescence (blue) spectral features, respectively. (b) RIXS spectrum collected using an excitation photon energy of 459.3 eV at an off-Bragg angle for the SW-RIXS measurements. The RIXS spectrum consists of a quasielastic, $d_{xz}/d_{yz}$, $d_{x^2-y^2}$, $d_{z^2}$, and fluorescence excitations. (c) Experimental (open circle) and best-fit (curve) rocking curves for $d_{xz}/d_{yz}$, fluorescence, $d_{x^2-y^2}$, and $d_{z^2}$ excitations. The color scheme in this panel adopts that in panel (b). (d) Determined depth profiles of these RIXS excitations in the top 5 interfaces of LAO/STO heterostructures.

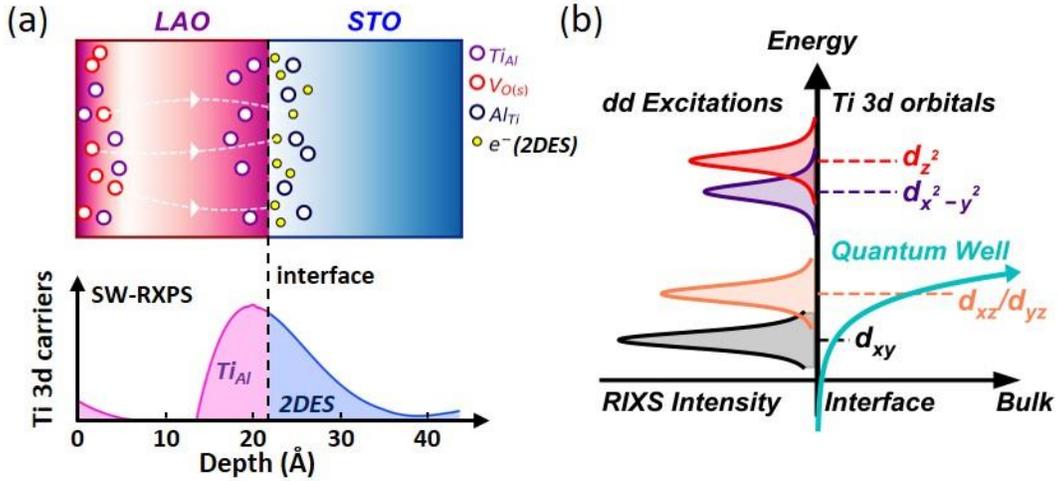

FIG. 4. Formation and subband energy levels of a 2DES at a LAO/STO interface. (a) Illustration of the polarity-induced defect mechanism that leads to the 2DES formation. Paired antisite defects ($Ti_{Al}$ in the LAO layer and $Al_{Ti}$ in the STO layer) form in order to alleviate the polarization-induced field across the LAO/STO interfaces. When the LAO layer is below the critical thickness (3 uc), all electrons transferred from $Ti_{Al}$ in the LAO layer are trapped by the deep $Al_{Ti}$ defects in the STO layer, causing no mobile electron. In contrast, with LAO layer thickness over 3 uc, the surface oxygen vacancy defects ($V_{O(S)}$) start to donate half of the electrons to $Al_{Ti}$ defects, and the rest to form the 2DES ($e^-$) in the STO layer. The $Ti_{Al}$ defects ($Ti^{3+}$) can be probed by RXPS and RIXS. The lower panel is the SW-RXPS-determined depth profile, showing two main Ti 3d sources. One source in STO is 2DES in the STO and the other is $Ti_{Al}$ in the LAO, matching the polarity-induced defect mechanism. (b) The orbital transitions from ground state ($d_{xy}$) to final states ($d_{xz}/d_{yz}$, $d_{x^2-y^2}$, and $d_{z^2}$) are responsible for the *dd* RIXS process. The distributions of these *dd* excitations are affected by the quantum confinement effect.

# Supplemental Materials

# Two-dimensional electron systems in perovskite oxide heterostructures: Role of the polarity-induced substitutional defects

Shih-Chieh Lin,[1,2] Cheng-Tai Kuo,[1,2,3,*] Yu-Cheng Shao,[4] Yi-De Chuang,[4] Jaap Geessinck,[5] Mark Huijben,[5] Jean-Pascal Rueff,[6,7] Ismael L. Graff,[8] Giuseppina Conti,[1,2] Yingying Peng,[9,#] Aaron Bostwick,[4] Eli Rotenberg,[4] Eric Gullikson,[2] Slavomír Nemšák,[4] Arturas Vailionis,[10,11] Nicolas Gauquelin,[5,12] Johan Verbeeck,[12] Giacomo Ghiringhelli,[9] Claus M. Schneider,[1,13] and Charles S. Fadley[1,2,†]

[1] *Department of Physics, University of California Davis, Davis, California 95616, USA*
[2] *Materials Sciences Division, Lawrence Berkeley National Laboratory, Berkeley, California 94720, USA*
[3] *Stanford Synchrotron Radiation Lightsource, SLAC National Accelerator Laboratory, Menlo Park, California 94025, USA*
[4] *Advanced Light Source, Lawrence Berkeley National Laboratory, Berkeley, California 94720, USA*
[5] *Faculty of Science and Technology and MESA+ Institute for Nanotechnology, University of Twente, Enschede 7500 AE, The Netherlands*
[6] *Synchrotron SOLEIL, L'Orme des Merisiers, Saint-Aubin-BP48, 91192 Gif-sur-Yvette, France*
[7] *Sorbonne Université, CNRS, Laboratoire de Chimie Physique-Matière et Rayonnement, 75005 Paris, France*
[8] *Department of Physics, Federal University of Paraná, Curitiba, Brazil*
[9] *CNR-SPIN and Dipartimento di Fisica Politecnico di Milano, Piazza Leonardo da Vinci 32, Milano I-20133, Italy*
[10] *Stanford Nano Shared Facilities, Stanford University, Stanford, California 94305, USA*
[11] *Department of Physics, Kaunas University of Technology, Studentu street 50, LT-51368 Kaunas, Lithuania*
[12] *Electron Microscopy for Materials Science (EMAT), University of Antwerp, Groenenborgerlaan 171, B-2020 Antwerp, Belgium*
[13] *Peter-Grünberg-Institut PGI-6, Forschungszentrum Jülich, Jülich 52425, Germany*
\* Email: ctkuo@slac.stanford.edu
# Present address: International Center for Quantum Materials, School of Physics, Peking University, Beijing 100871, China
† Deceased August 1, 2019.

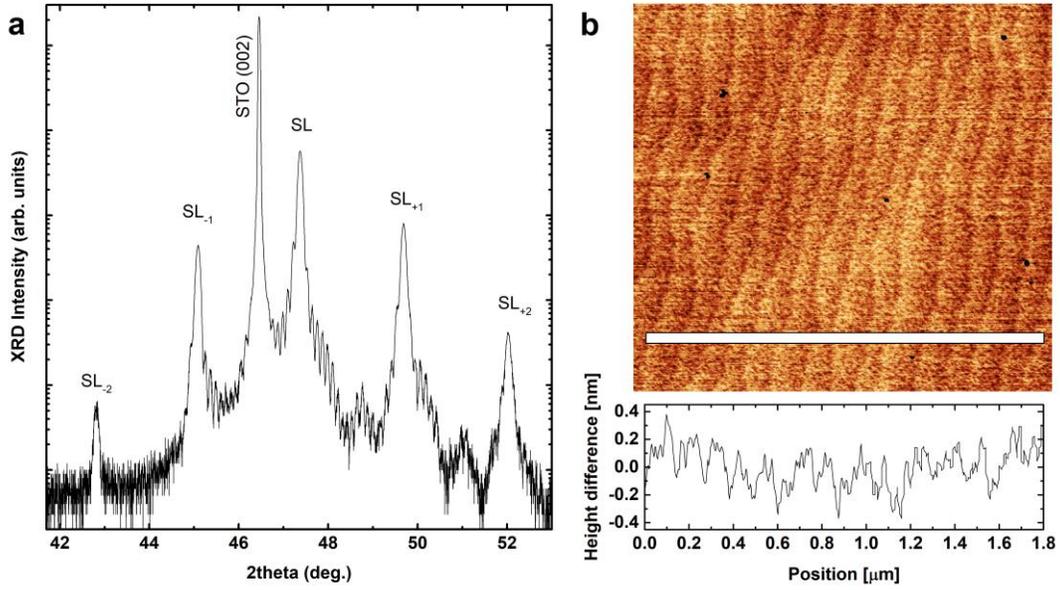

Supplemental FIG. 1. Structural characterization of the LAO/STO superlattice. (a) X-ray diffraction analysis of the out-of-plane crystalline ordering of the LAO/STO superlattice. The XRD results showed a reduced c-axis parameter of 3.73 Å for each LAO layer as compared to 3.90 Å for each STO layer, (b) Atomic force microscopy image (1.9 μm x 1.9 μm) of the superlattice surface. The topology and averaged line profile (white rectangle) show smooth terraces separated by clear, single-unit-cell height steps similar to the surface of the initial $TiO_2$-terminated STO (001) substrate.

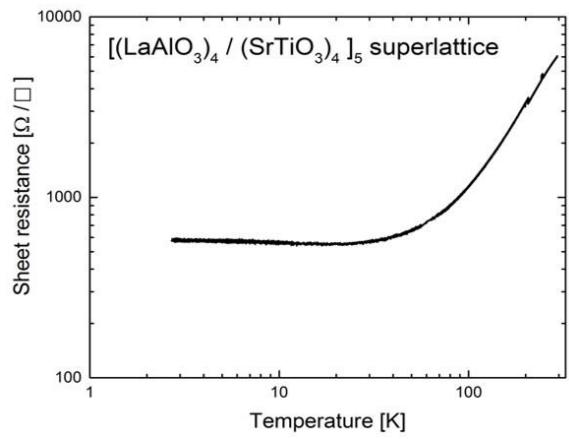

Supplemental FIG. 2. Metallic transport behavior of a $[(LaAlO_3)_4/(SrTiO_3)_4]_5$ superlattice.

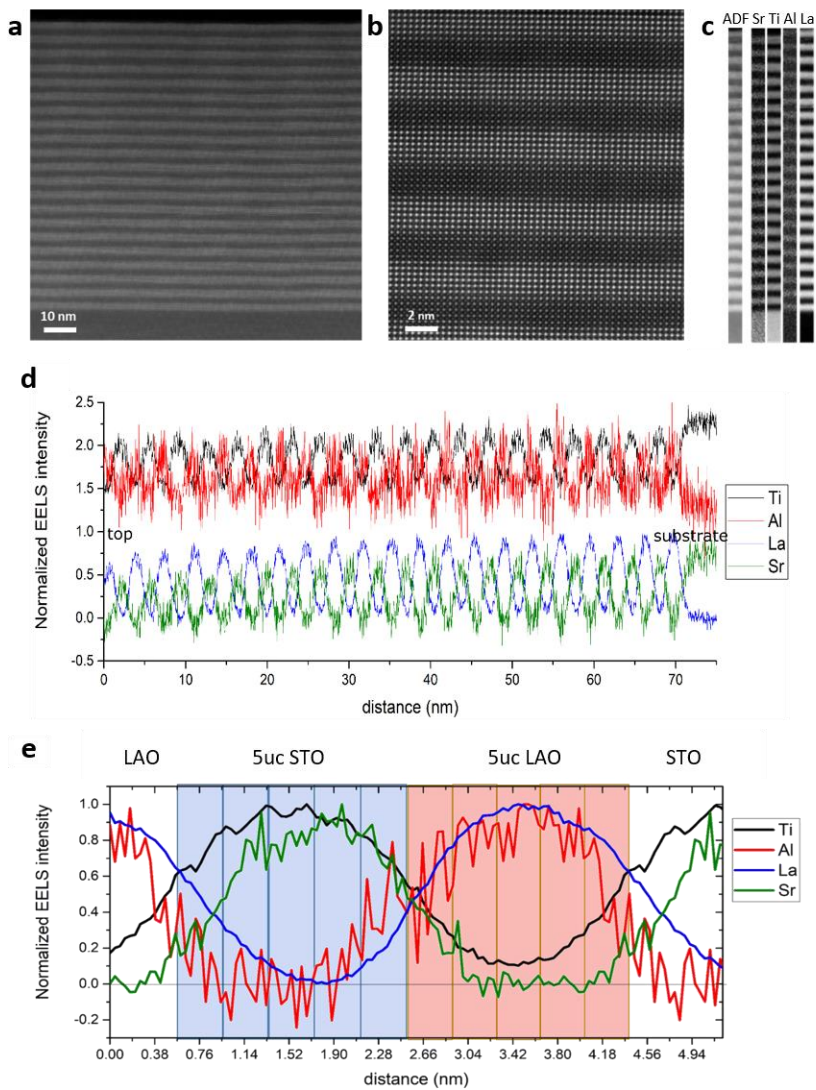

Supplemental FIG. 3. STEM characterisation of the LAO/STO superlattice. (a) HAADF overview image of the all sample of 20 repetitions of LAO(5uc)/STO(5uc) on STO substrate showing the great growth control of the specimen. (b) HAADF high magnification image of a region of 5,5 periods showing the presence of local roughness of +-1 uc at each interface. (c) EELS two-dimensional elemental maps of the Sr $L_{2,3}$, Ti $L_{2,3}$, Al K, and La $M_{4,5}$ edges with the corresponding ADF signal simultaneously acquired showing the great compositional control over each interfaces of the superlattice. (d) EELS line profile of the Ti $L_{2,3}$, Sr $L_{2,3}$, Al K and La $M_{4,5}$ edges showing the limited interdiffusion on both the A (Sr/La) and the B (Ti/Al) sites of the perovskite. (e) Average compositional profile over a region of 1,5 period of the superlattice showing the dissymmetry of the interfaces in terms of interdiffusion on the A site.

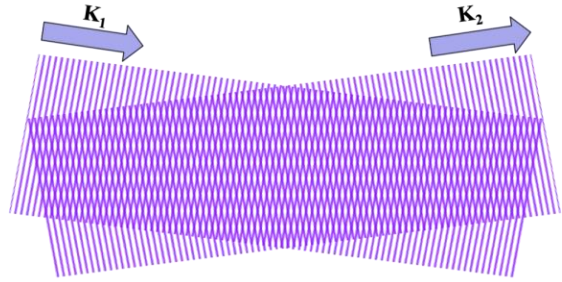

Supplemental FIG. 4. Schematic picture of a stationary wavefield formed by the interference of two coherent plane waves.

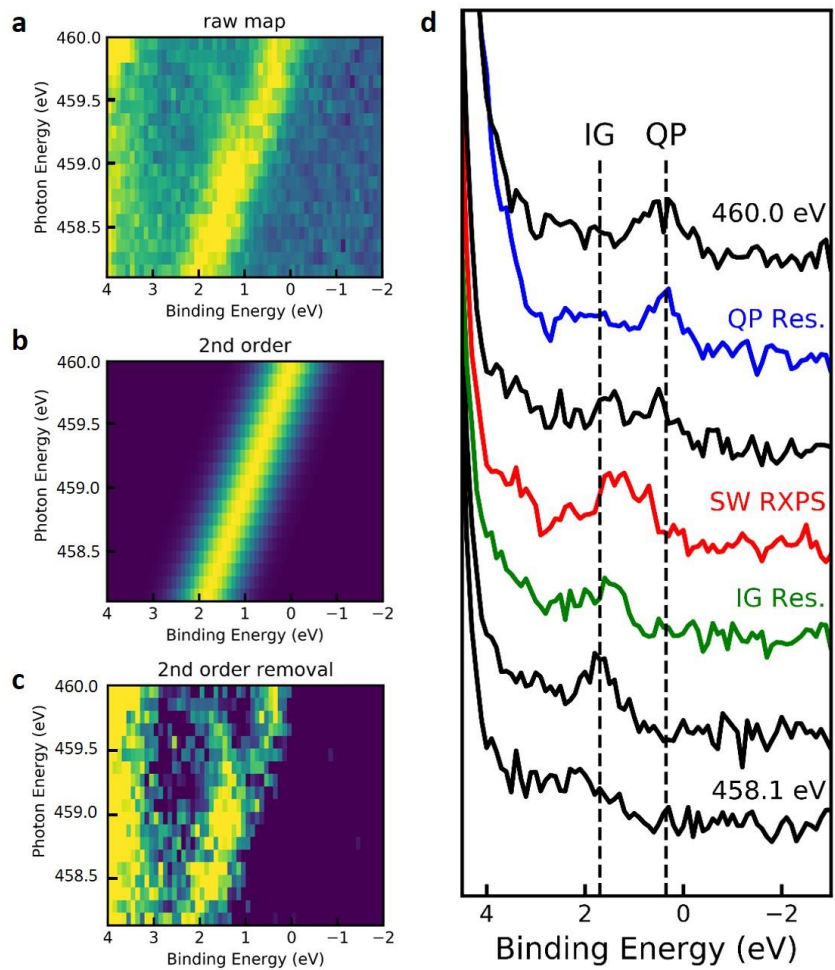

Supplemental FIG. 5. Photon-energy-scanned RXPS measurements for removal of 2nd order light contributions and identification of the contributions to the near Fermi (NF) peak of SW-RXPS measurements. (a) Raw map (b) 2nd order map (c) 2nd order removal map. (d) Corrected RXPS spectra at different resonances.

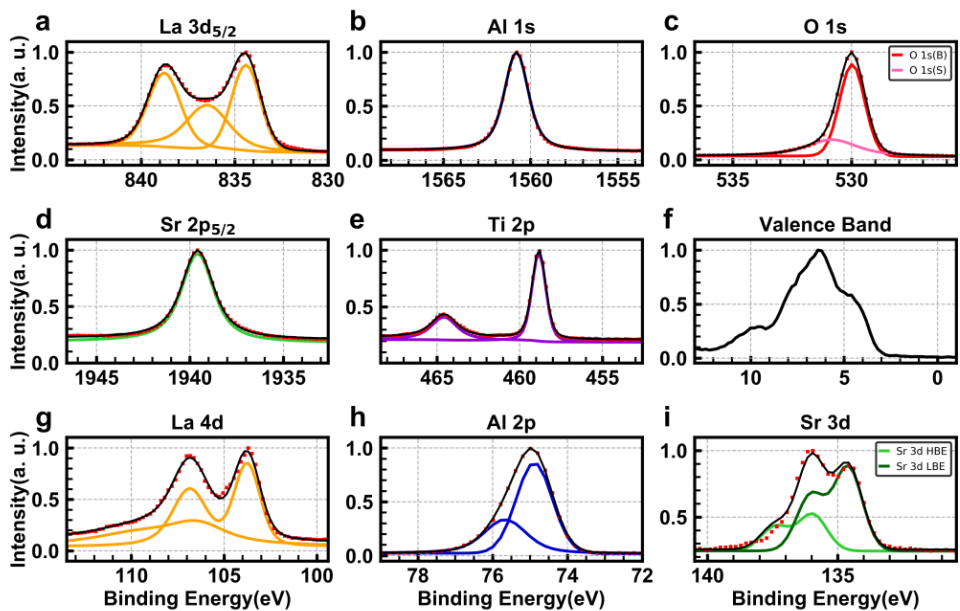

Supplemental FIG. 6. Core-level and valence band spectra collected by HXPS and RXPS. Core level and valence band spectra measured at an off-Bragg angle for (a)-(f) SW-HXPS and (g)-(i) SW-RXPS measurements. The core level spectra were fitted using Voigt function and Shirley background.

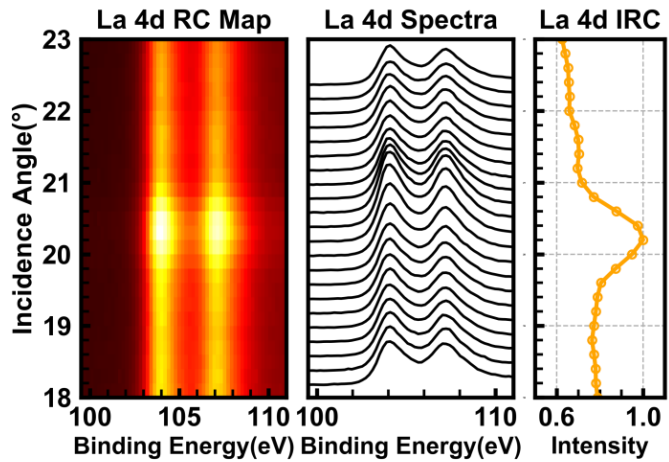

Supplemental FIG. 7. Demonstration of deriving an experimental RC. The left panel is the RC map of La 4*d* and the middle panel is the corresponding plots of La 4*d* spectra. Simply plot the integrated core-level peak intensities versus the incidence angles (integrated RC, IRC) can already show the intensity modulation of ~25%, a direct indication of SW effect. For the RCs in Fig. 2 and supplemental Fig. 8, The Shirley background contributions were removed for more careful investigations of the SW effect.

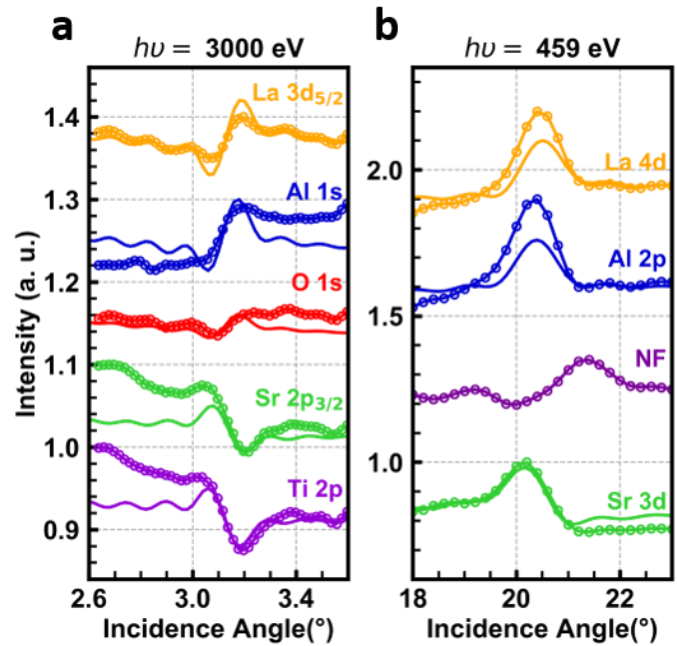

Supplemental FIG. 8. Core-level and NF rocking curves collected by SW-HXPS and SW-RXPS. (a) Core-level rocking curves collected by SW-HXPS. (b) Core-level and NF rocking curves collected by SW-RXPS.

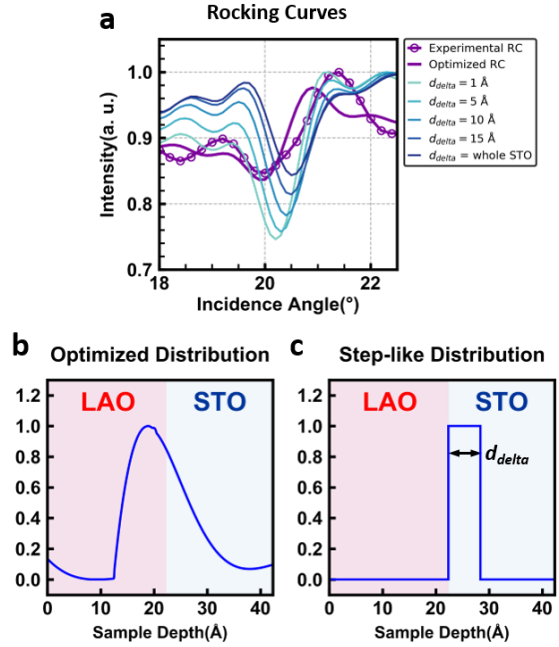

Supplemental FIG. 9. Comparison of the NF expt. RC with simulations using various NF depth distribution. (a) The NF expt. RC (also in Fig. 2(b)) and simulated RCs using optimized and step-like distributions. The optimized distribution is demonstrated in (b) and the step-like distributions is in (c). The best-fit RC shown in Fig. 2(b) was simulated using the optimized distribution. For the step-like distribution, when $d_{delta}$ = whole STO, it means that it is the case of uniform distribution in the whole STO layer. The errors of different distributions are summarized in supplemental Table 1.

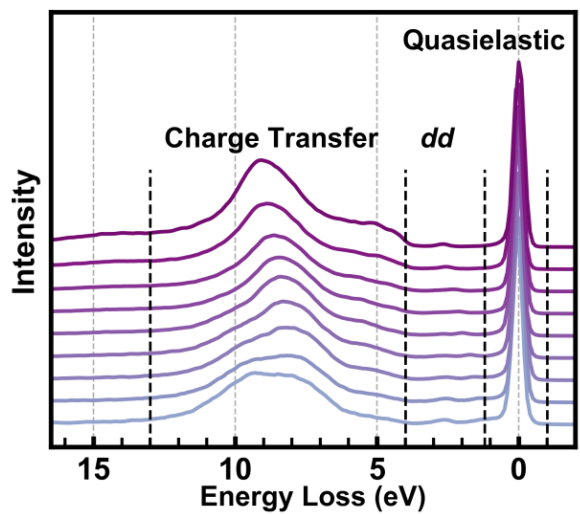

Supplemental FIG. 10. Excitation-energy-dependent RIXS spectra with wide energy loss range. These RIXS spectra were collected with identical excitation energy range and show the energy loss region of -2~16.5 eV. The quasielastic, *dd*, and charge transfer peaks are separated by the dashed lines.

Supplemental Table 1. Summary of mean square errors assuming different depth distribution

| Distribution | Mean square errors |
|---|---|
| **Optimized** | $9.1 \times 10^{-4}$ |
| **Step-like, $d_{delta}$ = 1 Å** | $3.1 \times 10^{-3}$ |
| **Step-like, $d_{delta}$ = 5 Å** | $2.8 \times 10^{-3}$ |
| **Step-like, $d_{delta}$ = 10 Å** | $2.8 \times 10^{-3}$ |
| **Step-like, $d_{delta}$ = 15 Å** | $2.9 \times 10^{-3}$ |
| **Step-like, $d_{delta}$ = whole STO** | $2.9 \times 10^{-3}$ |

**Supplemental Note 1**

**Characterization of the LAO/STO superlattice**

The superlattice sample, [(LAO)$_5$/(STO)$_5$]$_{20}$, was characterized to confirm exhibiting flat surface and high crystallinity. X-ray diffraction (XRD) data in Supplemental Fig. 1(a) show clear Laue fringes between the diffraction peaks. The atomic force microscopy image in Supplemental Fig. 1(b) shows smooth step terraces, similar to the surface of the initial TiO$_2$-terminated STO (001) substrate. The transport properties at the LAO/STO interface depend strongly on the LAO layer thickness when deposited on a STO substrate [3] and on the STO layer thickness [2,53]. A previous study [2] shows that adding an additional STO layer on top of the LAO layer triggers the electronic conductivity at a significantly lower LAO film thickness than for the uncapped systems, and even a STO capping layer of only 2 uc resulted in very similar metallic behavior as a 10 uc STO capping layer [7]. Those bilayer systems were realized on a STO single crystal substrate, which is known to result in interface conductivities with the highest electron mobilities in the order of $10^3$-$10^4$ cm$^2$V$^{-1}$s$^{-1}$ at ~10K (e.g. Refs [7]&[49] ), in strong contrast to LAO-STO bilayers deposited on other substrates, such as NdGaO$_3$ [54] for Huang et al. which only provides 10 cm$^2$V$^{-1}$s$^{-1}$ at ~10K. Therefore, for comparison between different studies the substrate has to be taken into consideration. Supplemental Fig. 2 shows an example of typical metallic behavior in a LAO/STO superlattice on a STO substrate with thicknesses of the individual LAO and STO layers of only 4 uc. In our superlattice the thicknesses of the LAO layers (5 uc) and STO layers (5 uc) are chosen such that the final structure will exhibit metallic behavior with high mobility charge carriers, similar to a single LAO/STO interface system.

Supplemental Fig. 3 shows the results of STEM characterization. On the one hand at the STO_bottom/LAO_top interface which can be assimilated to a "SrO/AlO$_2$/LaO" interface, Sr and Ti diffuse 1-1.5uc inside the LAO, and La and Al diffuse 1-1.5uc inside the STO, this results in the following stacking: TiO$_2$-SrO-Ti$_{0.6}$Al$_{0.4}$O$_2$-La$_{0.5}$Sr$_{0.5}$O-Ti$_{0.4}$Al$_{0.6}$O$_2$-La$_{0.9}$Sr$_{0.1}$O-Ti$_{0.1}$Al$_{0.9}$O$_2$-LaO. On the other hand, at the LAO_bottom/STO_top interface which can be assimilated to a "LaO/TiO$_2$/SrO" interface, Sr diffuse 1uc inside the LAO whereas Ti diffuses over 2uc , and Al diffuses 1uc inside the STO whereas La diffuses over 3 uc, this results in the following stacking: AlO$_2$-LaO-Ti$_{0.2}$Al$_{0.8}$O$_2$-La$_{0.9}$Sr$_{0.1}$O-Ti$_{0.7}$Al$_{0.3}$O$_2$-

La$_{0.5}$Sr$_{0.5}$O-TiO$_2$-La$_{0.2}$Sr$_{0.8}$O-TiO$_2$-La$_{0.1}$Sr$_{0.9}$O. Note that a diffusion of 1uc can be attributed to roughness and the accuracy of concentrations can be overestimated to 0.1. We can therefore conclude that the STO_bottom/LAO_top interface is sharper whereas the LAO_bottom/STO_top interface is more diffused. This confirms observation made by other authors [12,50] and theoretical predictions [16]. The presence of around 10% of Ti at the center of the LAO indicating a larger diffusion length of Ti inside the LAO (of the order of 2.5-3uc) cannot be excluded but can be mitigated by the presence of electron delocalization/channeling over 4-5 Angstrom (1uc) beyond the scope of this paper.

**Supplemental Note 2**

**Basic concepts of standing wave techniques**

The X-ray standing wave (SW) technique is an interference method that provides excellent depth resolution of atomic and electronic structure of materials. The SW technique uses constructive/destructive interference between the incident and reflected X-ray beam to tailor the intensity profile of the electric field along the sample depth. In general, the yield of an X-ray probing technique is proportional to the strength of electric filed. Hence, by applying SW technique, one can selectively enhance or suppress the yield from a specific depth of the sample, such as a buried layer or an interface. Moreover, since the wavelengths of the excited X-ray is in (sub-) nm range, the spatial resolution of a SW technique is as high as few angstroms [55, 56].

**Formation of the standing wavefield**

As shown in Supplemental Fig. 4, a stationary wavefield is created by the interference between the incident and Bragg-reflected X-rays. Note that to form a stationary wavefield, a Bragg diffraction need to be utilized to enhance the X-ray reflectivity for creating sufficient strength of interference. The incident and reflected X-rays can be described by the following equations

$$\mathbf{E_1} = \boldsymbol{e_1} E_1 e^{i(\omega_1 t - \mathbf{K_1 r})} \quad \text{and} \quad \mathbf{E_2} = \boldsymbol{e_2} E_2 e^{i(\omega_2 t - \mathbf{K_2 r})} \qquad \text{Eq. S.1}$$

where **E₁** and **E₂** are the electric field vectors, $e_1$ and $e_2$ are the polarization vectors of the incident and reflected wave, and the **K₁** and **K₁** are the X-ray propagation vectors. Then we represent the **E₂** by the following equation

$$E_2 = \sqrt{R}E_1 \exp(i\vartheta) \qquad \text{Eq. S.2}$$

where the $R$ is the reflectivity and $\vartheta$ is a phase factor. By substituting the $E_2$ in Eq. S.1 with the RHS of Eq. S.3, the electric field **E** of the SW is derived by the superposition of incident and reflected X-ray.

$$\mathbf{E} = \mathbf{E_1} + \mathbf{E_2} = e^{i\omega t}\left[e_1 E_1 e^{(-i\mathbf{K_1 r})} + e_2 \sqrt{R} E_1 e^{i\vartheta} e^{i(-i\mathbf{K_2 r})}\right] \qquad \text{Eq. S.3}$$

Then, the wavefield intensity $I$ is simply derived from the strength of electric field **E** by applying cosine theorem.

$$I = |\mathbf{E}|^2 = I_1(1 + R + 2\sqrt{R}\cos(\vartheta - \mathbf{Hr})) \qquad \text{Eq. S.4}$$

where **H** = **K₂** - **K₁** and $\vartheta$ is the phase difference between incident and reflected X-ray. Moreover, since the stationary wavefield is formed by utilizing Bragg reflection, the wavelength of this wavefield is equal to the planar spacing $d_H$ in the Bragg equation

$$2d_H \sin\theta = n\lambda \qquad \text{Eq. S.5}$$

Note that the $\lambda$ in Eq. S.5 is the wavelength of incident and reflected waves. In this study, the planar spacing $d_H$ is the thickness of a repeating unit of LAO/STO superlattice which is 39.2 Å.

Furthermore, given that the probability for the concomitant emission from A atom is proportional to the intensity of the electric field at the center of A, the yield is described by the following equation

$$Y_A^H = Y_{A,1}(1 + R + 2\sqrt{R}\cos(\vartheta - \mathbf{Hr_A})) \qquad \text{Eq. S.5}$$

where $Y_{A,1}$ is the yield with no SW effect and $\mathbf{r_A}$ is the location of this atom.

**Depth resolution**

For a SW experiment, the capability of offering depth resolution is rooted in two characteristics: capability to distinguish the yields from different depths and the tunability of the position of SW wavefield. The capability to differentiate the yields from different depth is determined by the magnitude of intensity modulation of the interference. The maximum of intensity modulation occurs when the reflectivity R equals to 1. In this case, the intensity profile is consisting of alternating completely dark, $I = 0$, and light, $I = 4 I_0$, region which results in the greatest intensity modulation, $4I_0$.

The position of SW wavefield can be controlled by the varying the phase factor $\vartheta$ in Eq. S.5. The SW wavefield moves while the phase $\vartheta$ changes. Moreover, while sweeping over the entire range of a Bragg reflection, the phase factor changes by $\pi$ and the SW traverses half of the planar spacing, $d_H$. In this experiments, the wavefield is moved by varying the incidence angle. Finally, since the depth resolution of SW techniques is dominated by their modulated interference pattern, it is intuitive to define the depth precision as a fraction of the wavelength of SW wavefield. A general rule of thumb for the depth precision for a SW measurement is $0.1\times \lambda_{SW}$, which is around 2-4 Å [39]. In this study, since $\lambda_{SW} \approx d_H = 39.2$ Å, the depth precision is ~3.92 Å. The error of spatial resolution caused by the photon energy resolution can be negligible for most cases. Taking our study as an example, considering the energy resolution 0.5 eV, Bragg angle of 20 deg, and photon energy of 459 eV, the error caused by the energy resolution is 0.043 Å based on the following calculation:

$$error_{\delta E} = \frac{\partial (d_H)}{\partial E} dE = \frac{\partial (\lambda/2\sin\theta)}{\partial E} dE, \text{ with } \lambda(\text{Å}) = \frac{12400}{E(eV)}$$

$$= \frac{12400}{2\sin\theta} \frac{\partial (1/E)}{\partial E} dE = -\frac{12400}{0.684} \frac{1}{459^2} * 0.5 = 0.043 \text{ Å}$$

This error is much smaller comparing to the precision limited by the wavelength of SW wavefield (3.92 Å). Hence, we conclude that the spatial error caused by photon energy resolution is negligible in this experiment.

**Rocking Curves**

As presented in Fig. 2(a), 2(b) and 3(c), rocking curves are the primary type of data representation of SW experiments from which the depth profiles of electronic or atomic structures are acquired. As discussed before, the X-ray standing wavefield is shifting by half of the planar spacing $d_H$ while changing the incidence angle through the Bragg condition. As the wavefield shifts through the position of a specific atom, the yield from this site is modulated, and this intensity modulation is called a rocking curve. In this study, the strength of yield are the intensities of corresponding photoemission core-level and valence states and RIXS excitations determining by curve fittings method.

Note that in the range of a period of the standing wavefield, the yield from every depth of the sample has a unique rocking curve. Based on this fact, we can extract the depth profile of the atomic or electronic structure by implementing an X-ray simulation code, Yang's X-ray Optics (YXRO) [39].

**Supplemental Note 3**
**Removal of 2$^{nd}$ order contributions and identification of the near Fermi (NF) peak in the RXPS spectra**

Supplemental Fig. 5(a)-(c) demonstrates how the 2$^{nd}$ order contributions were removed. Supplemental Fig. 5(a), (b), and (c) are raw, 2$^{nd}$ order light, and 2$^{nd}$ order removal intensity map, respectively. The 2$^{nd}$ order peaks are fitted using the Gaussian lineshape observed at higher photon energies. The 2$^{nd}$ order peak has 2Δhν energy shift if the photon energy changes by Δhν. The linear trend of the NF peak in the raw map is due to the influence of this 2$^{nd}$ order peak. After its removal, one can see that the linear trend is gone in the removal intensity. Supplemental Fig. 5(c) shows a corrected photon-energy-scanned RXPS VB map scanned with the incoming photon energies across Ti 2p. A closer look of the evolution of the quasiparticle states (QP) and in-gap (IG) peaks can be done by seeing the peak shape of these spectra at various resonance, as shown in Supplemental Fig. 5(d). One can find that at the IG resonance, the IG peak at ~1.5 eV has significant intensity, while at the QP resonance the QP peak at ~0.2 eV is most significant. Comparing to the literature [24,45], our results agree well with this trend. However, one can notice the limited energy resolution in our data: The FWHM of the QP peak in this work is ~ 1eV, which is wider than that of the prior work [24]. This makes the QP and IG peaks not distinguishable at the chosen energy for SW-RXPS measurements. This specific energy was chosen for maximizing the reflectivity/standing wave effect but apparently it causes the drawback of indistinguishable IG and QP contributions. We can conclude that the NF peak in the inset of Fig. 2(c) includes both contributions from QS and IG states.

**Supplemental Note 4**

**Fitting of core-level rocking curves**

A core-level rocking curve (RC) is defined as the fitted core-level peak intensities versus incidence angles. Supplemental Fig. 6 shows the strongest core-level spectra for all atomic species in the LAO/STO superlattice at an off-Bragg angle. The core levels were fitted using Shirley background and Voigt function. The demonstration of observing an SW effect in the raw data is shown in Supplemental Fig. 7. The left panel in Supplemental Fig. 7 is the La 4$d$ RC map and the middle panel shows the corresponding core-level spectra at a series of incidence angles. By simply integrating the peak intensity along the binding energy axis at each incidence angle (so called IRC), one can observe clear intensity modulation of 25%, as shown in the right panel of Supplemental Fig. 7. To have more careful investigation of SW effects, we use the fitted core-level peak intensities (so called RC) for all the analyses in this work. The experimental and best-fit RCs are demonstrated in Supplemental Fig. 8.

**Supplemental Note 5**

**Comparison of simulated RCs using various depth distribution**

We use the NF RC as an example to show that the difference between the experimental and simulated RCs using the optimized (best-fit), step-like ($d_{delta}$=1~15 Å), and uniform ($d_{delta}$= whole STO layer), distributions. Supplemental Fig. 9(a) shows the NF experimental RC along with simulated RCs using various depth distribution. Simply inspect by eyes, the simulated RC using optimized distribution is close to the experimental data the most. Supplemental Figs. 9(b)&(c) shows the optimized distribution and step-like distribution. The step-like distribution means that we assume the weighted factor (or concentration) equals one when $d = 0$ to $d_{delta}$ in the STO layer and the weighted factor equals zero elsewhere. In our analyses, the determined distributions for the best-fit results depend on the mean square errors. Supplemental table 1 summarizes the mean square errors of various distribution for the NF RC. It shows that the optimized distribution has the minimum errors ($9.1 \times 10^{-4}$) while other distributions have roughly errors of $\sim 3 \times 10^{-3}$. Therefore, the optimized distribution should be close to the reality the most.

Since the experimental NF RC does not match either the RC of a single LAO (STO) layer nor to the RC of a step-like layer, it suggests that the depth distribution of the NF peak has a density profile. In other words, it is a weighting coefficient function $W(z)$ that has various weighting coefficients at specific depths. The experimental NF RC has been fit by the equation:

$$I_{RC}^{Expt}(\theta) = \sum_{z_i} W(z_i) I_{RC}^{Calc}(z_i, \theta) \times \exp(-z_i / \Lambda_{x,eff})$$

where $I_{RC}^{Expt}(\theta)$ is an experimental RC at an incidence angle $\theta$, and $I_{RC}^{Calc}(z_i, \theta) \times \exp(-z_i / \Lambda_{x,eff})$ is the calculated RC from each delta layer, and $W(z_i)$ is a weighting coefficient at a depth $z_i$. The delta-layer RC, $I_{RC}^{Calc}(z_i, \theta)$, multiplied by a attenuation term $\exp(-z_i / \Lambda_{x,eff})$, has been calculated via the X-ray optics computing code (Yang X-ray Optics, YXRO). The depth profile $W(z)$ is thus determined in the fitting procedure that we have derived using the Broyden, Fletcher, Goldfarb, and Shanno (BFGS) method [42]. Finally, the summed amplitudes of these weighting factors at a given $z$, which yields a quantitative estimate of the depth distributions, is shown in Fig. 2(d).

**Supplemental Note 6**
**Wide-energy-loss-range RIXS spectra**

Supplemental Fig. 10 shows the excitation-energy-dependent RIXS spectra with energy los region (-2~16.5 eV).